\documentclass[iop]{emulateapj}

\usepackage{natbib,graphicx,amsmath}

\shorttitle{POST-NEWTONIAN APPROXIMATION VS. PERTURBATION THEORY}
\shortauthors{NOH \& HWANG}

\newcommand{\bea}{\begin{eqnarray}}
\newcommand{\eea}{\end{eqnarray}}

\begin{document}

\title{Cosmological post-Newtonian approximation compared with perturbation theory}
\author{Hyerim Noh${}^{1}$ and Jai-chan Hwang${}^{2,3}$}
\address{${}^{1}$Korea Astronomy and Space Science Institute, Daejon 305-348,
                Republic of Korea \\
         ${}^{2}$Department of Astronomy and Atmospheric Sciences,
                Kyungpook National University, Daegu 702-701, Republic of Korea \\
         ${}^{3}$Korea Institute for Advanced Study, Seoul 130-722, Republic of
                Korea}


\begin{abstract}
We compare the cosmological first-order post-Newtonian (1PN) approximation with the relativistic cosmological linear perturbation theory in a zero-pressure medium with the cosmological constant. We compare equations and solutions in several different gauge conditions available in both methods. In the PN method we have perturbation equations for density, velocity and gravitational potential independently of the gauge condition to 1PN order. However, correspondences with these 1PN equations are available only in certain gauge conditions in the perturbation theory. Equations of perturbed velocity and the perturbed gravitational potential in the zero-shear gauge exactly coincide with the Newtonian equations which remain valid even to 1PN order (the same is true for perturbed velocity in the comoving gauge), and equations of perturbed density in the zero-shear gauge and the uniform-expansion gauge coincide to 1PN order. We identify other correspondences available in different gauge conditions of the perturbation theory.
\end{abstract}


%
%
\section{Introduction}

Einstein's gravity is generally accepted as the gravity to handle astronomical phenomena. The theory holds a remarkable track record in the solar-system test based on vacuum Schwarzschild solution and the parameterized post-Newtonian approximation where the gravitational fields are supposed to be weak. It is true that Einstein's theory has not failed in any experimental test based on modern scientific and technological development up till today, but it is also true that there has been no experimental test of the theory in the strong gravitational field and in large scale even including the galactic scale. Einstein's gravity is generally accepted in cosmology mainly based on its successes in other astronomical and Earth bound tests and the theory's own prestige associated with Einstein's fame and historical legacy.

A self-consistent treatment of cosmological world model is possible in Einstein's gravity. Without a lead by Einstein's gravity, however, the spatially homogeneous and isotropic cosmological world model based on Newton's gravity is known to be incomplete and indeterminate (Layzer 1954; Lemons 1988). Despite such troubles in the background world model, evolution of perturbations in Newton's theory is known to be quite successful in reproducing the corresponding results in Einstein's gravity (Lifshitz 1946; Bonnor 1957; Noh \& Hwang 2004; Jeong et al. 2011). Considering the action-at-a-distance nature of Newton's gravity, such a coincidence is a non-trivial result. Differences between the two theories, however, appear as the scale approaches the horizon. We will address this issue in this work.

If we accept Einstein's theory in analyzing the large-scale cosmic structure in current era, we have two methods available. One well known method is the perturbation theory where all dimensionless deviations in the metric and the energy-momentum tensor from the background world model are assumed to be small. If we accept only linear order deviations we have the linear perturbation theory. The perturbation theory assumes small deviation but is fully relativistic, generally applicable in all scales including the super-horizon scale and to particles with relativistic velocities (Lifshitz 1946; Harrison 1967; Nariai 1969; Bardeen 1980; Peebles 1980; Kodama \& Sasaki 1984; Bardeen 1988; Mukhanov et al. 1992; Ma \& Bertschinger 1995).

The other less known method is the cosmological post-Newtonian (PN) approximation where all dimensionless deviations are assumed to be weakly relativistic with ${GM \over Rc^2} \ll 1$ ($M$ and $R$ are characteristic mass and length scales) and for a virialized system ${v^2 \over c^2} \ll 1$ ($v$ is the characteristic velocity involved). The first-order PN (1PN) approximation makes expansion up to ${GM \over Rc^2} \sim {v^2 \over c^2}$ order. The PN approximation assumes small relativistic effects and is applicable only in sub-horizon scale. But the equations derived are applicable to fully nonlinear situations (Chandrasekhar 1965; Futamase 1988; Tomita 1988; Futamase 1989; 
Tomita 1991; Futamase 1993a, 1993b; Shibata \& Asada 1995; Asada \& Futamase 1997; Tanaka \& Futamase 1999; Hwang et al. 2008, PN2008 hereafter). Therefore, the two methods are complementary to each other.

If we encounter cosmological situations where both nonlinearity as well as relativistic effects are important we may need full-blown numerical relativity implemented in cosmology. Currently such a general relativistic numerical simulation in cosmology is not available. The nonlinear perturbation analysis, being based on perturbative approach, is not sufficient to handle the genuine nonlinear aspects of structure formation accompanied with self-organization and spontaneous formation of structures. In order to handle the relativistic nonlinear process in cosmology we believe the post-Newtonian approach is currently practically relevant to implement in numerical simulation.

We can find cosmological situations where the cosmological post-Newtonian approach, being weakly relativistic but fully nonlinear, might have important applications. Especially, the current cosmological paradigm favors a model where the large-scale structures (requiring the relativistic treatment) are in the linear stage, whereas small-scale structures are apparently in fully nonlinear stage. The often adopted strategy is to assume the small-scale nonlinear structures are fully under control by the Newtonian gravity. In the galactic and cluster scales we have the general relativistic measure ${GM \over Rc^2} \sim {v^2 \over c^2} \sim 10^{-6} - 10^{-4}$, thus small but nonvanishing, and indeed the 1PN (weakly relativistic) assumption is quite sufficiently valid. Thus, we believe the 1PN approach would be quite relevant to estimate the general relativistic effects in the nonlinear clustering processes of the galaxy cluster-scale and the large-scale structures.

In this work we will compare the two relativistic methods in the matter dominated era: the 1PN method vs. linear perturbation theory. The 1PN approximation is based on previous studies in Chandrasekhar (1965) and PN2008, whereas the linear perturbation theory is based on previous studies in Bardeen (1988), Hwang (1994), and Hwang \& Noh (1999). We will compare the equations and solutions derived in the two methods. In both methods we have gauge degrees of freedom which need to be fixed by the gauge conditions. The Newtonian perturbation theory appears as the zeroth-order PN (0PN) approximation. Thus, we naturally also have Newtonian theory for the comparison. To 1PN order we will show that the equations for density, velocity and gravitational potential do not depend on the gauge conditions with each variables gauge-invariant to 1PN order. In the perturbation theory, however, the perturbation variables for density, velocity, potential, curvature, and other kinematic variables (like expansion and shear) do depend on the gauge conditions adopted. Our emphasis in this work is on the correspondences between the PN variables and perturbation theory variables based on different gauge conditions.

For the velocity and gravitational potential the Newtonian (0PN) equations are valid even to 1PN order, whereas for the density variable we have 1PN correction terms: see Equations (\ref{PN-delta-eq1})-(\ref{PN-U-eq}).
The zero-shear gauge will be shown to be distinguished by showing the perturbed velocity and gravitational potential having exact correspondences with the 0PN (thus 1PN as well) equations, and the perturbed density showing correct correspondence with the 1PN equation. The zero-shear gauge can be contrasted with the comoving gauge where the perturbed density and velocity have exact Newtonian correspondences even to the second order perturbations in all scales whereas the gravitational potential vanishes to the linear order. Other PN correspondences of the perturbation theory are summarized in Section \ref{sec:correspondences}.

%
%
\section{1PN approximation}

To the 1PN order our metric convention is (Chandrasekhar 1965; Chandrasekhar \& Nutku 1969; PN2008) \bea
   & & \!\!\!\!\!\!\!\!\!\!
       ds^2 = - \left[ 1 - {1 \over c^2} 2 U
       + {1 \over c^4} \left( 2 U^2 - 4 \Phi \right) \right] c^2 d t^2
   \nonumber \\
   & & \!\!\!\!\!\!\!\!\!\!
       \qquad
       - {1 \over c^3} 2 a P_i c dt d x^i
       + a^2 \left( 1 + {1 \over c^2} 2 V \right) \gamma_{ij} d x^i d x^j,
   \label{metric-1PN} \\
   \nonumber
\eea where $a(t)$ is a cosmic scale factor. We consider a {\it flat} background, thus $\gamma_{ij} = \delta_{ij}$. Then, we have $V = U$ (PN2008).
By ignoring $c^{-2} a^2 C_{ij}$ part in the $g_{ij}$ we already have taken spatial gauge conditions without losing any generality: see Section 6 in PN2008.

\subsection{Equations in the zero-pressure case}

We consider a zero-pressure fluid in the presence of cosmological constant.
The 1PN energy-momentum tensor is presented in Equation (21) of PN2008 (Chandrasekhar 1965).
With vanishing pressure, internal energy, flux, and anisotropic stress, we have \bea
   & & T_{ab} = \widetilde \mu \widetilde u_a \widetilde u_b, \quad
       \widetilde \mu \equiv \varrho c^2, \quad
       \widetilde u^i \equiv {1 \over c} {1 \over a} v^i \widetilde u^0,
\eea where $\widetilde \mu$ and $\widetilde u_a$ are the covariant energy density and the normalized fluid four-vector, respectively.
To the linear order, we have (PN2008)
\bea
   T^0_0 = - \varrho c^2, \quad
       T^0_i = a \varrho c \left( v_i - {1 \over c^2} P_i \right),
       \quad
       T^i_j = 0.
   \label{Tab-1PN}
\eea The 1PN equations are (Chandrasekhar 1965; PN2008) \begin{widetext} \bea
   & & \!\!\!\!\!\!\!\!\!\!
       {1 \over a^3} \left( a^3 \varrho \right)^{\displaystyle\cdot}
       + {1 \over a} \left( \varrho v^i \right)_{|i}
       = - {1 \over c^2} \varrho \left( {\partial \over \partial t}
       + {1 \over a} {\bf v} \cdot \nabla \right)
       \left( {1 \over 2} v^2 + 3 U \right),
   \label{mass-conservation} \\
   & & \!\!\!\!\!\!\!\!\!\!
       {1 \over a} \left( a v_i \right)^{\displaystyle\cdot}
       + {1 \over a} v_{i|j} v^j
       - {1 \over a} U_{,i}
       = {1 \over c^2} \left[
       {1 \over a} v^2 U_{,i}
       + {2 \over a} \left( \Phi - U^2 \right)_{,i}
       + {1 \over a} \left( a P_i \right)^{\displaystyle\cdot}
       + {2 \over a} v^j P_{[i|j]}
       - v_i \left( {\partial \over \partial t}
       + {1 \over a} {\bf v} \cdot \nabla \right)
       \left( {1 \over 2} v^2 + 3 U
       \right)
       \right],
   \label{momentum-conservation} \\
   & & \!\!\!\!\!\!\!\!\!\!
       {\Delta \over a^2} U
       + 4 \pi G \left( \varrho - \varrho_b \right)
       = - {1 \over c^2} \left\{
       {1 \over a^2} \left[
       2 \Delta \Phi
       - 2 U \Delta U
       + \left( a P^i_{\;\;|i} \right)^{\displaystyle\cdot} \right]
       + 3 \ddot U
       + 9 {\dot a \over a} \dot U
       + 6 {\ddot a \over a} U
       + 8 \pi G \varrho v^2
       \right\},
   \label{Raychaudhury-eq} \\
   & & \!\!\!\!\!\!\!\!\!\!
       0 = - {\Delta \over a^2} P_i
       - 16 \pi G \varrho v_i
       + {1 \over a} \left( {1 \over a} P^j_{\;\;|j}
       + 4 \dot U + 4 {\dot a \over a} U \right)_{,i},
   \label{momentum-constraint}
\eea
\end{widetext}
where a vertical bar indicates spatial covariant derivative based on $\gamma_{ij}$ as the metric.
These follow from the mass-conservation, momentum-conservation, Raychaudhury ($G^0_0 - G^i_i$), and momentum-constraint ($G^0_i$) equations, respectively; for the general case, see Equations (114), (115), (119) and (120) in PN2008. Terms on the left-hand-side provide the Newtonian (0PN) limit, and the ones in the right-hand-side are 1PN contributions. Notice that the 1PN terms include up to fourth-order perturbations.

To perturbed order we have $\varrho = \varrho_b + \delta \varrho$ where we will ignore the subindex $b$ indicating the background quantity unless necessary.

Our cosmological PN approach {\it assumes} near flat background, but is valid in the presence of the cosmological constant. Equations for the background \bea
   & & {\ddot a \over a}
       = - {4 \pi G \over 3} \varrho
       + {\Lambda c^2 \over 3}, \quad
       {\dot a^2 \over a^2}
       = {8 \pi G \over 3} \varrho
       + {\Lambda c^2 \over 3},
   \nonumber \\
   & &
       \dot \varrho + 3 {\dot a \over a} \varrho = 0,
   \label{BG-eqs}
\eea were {\it subtracted} in deriving the PN equations; see Section 3.2 in PN2008.

\subsection{Newtonian equations as the 0PN limit}

The 0PN approximation gives the Newtonian limit. The 0PN metric is
\bea
       ds^2 = - \left( 1 - {1 \over c^2} 2 U \right) c^2 d t^2
       + a^2 \delta_{ij} d x^i d x^j.
   \label{metric-0PN}
\eea To 0PN limit Equations (\ref{mass-conservation})-(\ref{Raychaudhury-eq}) give
\bea
   & & \dot \delta = - {1 \over a} \nabla \cdot
       \left[ \left( 1 + \delta \right) {\bf v} \right],
   \label{Newtonian-eq1} \\
   & & \dot {\bf v} + {\dot a \over a} {\bf v}
       + {1 \over a} {\bf v} \cdot \nabla {\bf v}
       = {1 \over a} \nabla U,
   \label{Newtonian-eq2} \\
   & & - {\Delta \over a^2} U = 4 \pi G \varrho \delta,
   \label{Newtonian-eq3}
\eea where $\delta \equiv \delta \varrho/\varrho$, ${\bf v} \equiv v^i$, and $\delta \Phi \equiv -U$ are the relative density contrast, velocity perturbation, and the perturbed Newtonian gravitational potential, respectively.

In the Newtonian context, {\it assuming} the presence of spatially homogeneous and isotropic background the above equations follow from the mass conservation, the momentum conservation, and the Poisson's equation, respectively (Peebles 1980); for  weakness of Newton's theory in handling background world model, however, see Layzer (1954) and Lemons (1988). In fact, the complete nonlinear Newtonian hydrodynamic equations are already built in as the 0PN approximation of Einstein's gravity (Chandrasekhar 1965; Bertschinger \& Hamilton 1994; Kofman \& Pogosyan 1995; see Section 4.2 in PN2008). Up to this point is a review of the cosmological 1PN approach.

To the linear order, from Equations (\ref{Newtonian-eq1})-(\ref{Newtonian-eq3}) we can derive \bea
   & & \!\!\!\!\!\!\!\!\!\!
       \ddot \delta + 2 H \dot \delta - 4 \pi G \varrho \delta
       = {1 \over a^2 H} \left[ a^2 H^2 \left( {\delta \over H}
       \right)^{\displaystyle\cdot} \right]^{\displaystyle\cdot}
       = 0,
   \label{delta-eq-N} \\
   & & \!\!\!\!\!\!\!\!\!\!
       \left( a \nabla \cdot {\bf v} \right)^{\displaystyle\cdot\displaystyle\cdot}
       + H \left( a \nabla \cdot {\bf v} \right)^{\displaystyle\cdot}
       - 4 \pi G \varrho
       \left( a \nabla \cdot {\bf v} \right) = 0,
   \label{v-eq-N} \\
   & & \!\!\!\!\!\!\!\!\!\!
       \left( a U \right)^{\displaystyle\cdot\displaystyle\cdot}
       + 2 H \left( a U \right)^{\displaystyle\cdot}
       - 4 \pi G \varrho
       \left( a U \right)
       = 0,
   \label{delta-Phi-eq-N}
\eea where $H \equiv \dot a/a$. {\it Assuming} $K = 0 = \Lambda$, the growing mode solutions are \bea
   & & \!\!\!\!\!
       \delta = - {2 \over 5} {c^2 \Delta \over a^2 H^2} C, \quad
       {1 \over a} \nabla \cdot {\bf v}
       = {2 \over 5} {c^2 \Delta \over a^2 H} C, \quad
       {1 \over c^2} U = {3 \over 5} C,
   \nonumber \\
   \label{sol-N}
\eea where $C({\bf x})$ is an integration constant; the normalization of the constant $C$ will become clear later, see below Equation (\ref{varphi_v-eq}). Notice that as the Newtonian (0PN) equations are not aware of the presence of horizon, Newtonian equations are supposed be valid in {\it all} scales. The Newtonian or the 0PN order equations are not affected by the gauge transformation.

\subsection{Linear limit of 1PN equations}

By keeping only linear order terms Equations (\ref{mass-conservation})-(\ref{momentum-constraint}) give \bea
   & & {1 \over a^3} \left( a^3 \delta \varrho \right)^{\displaystyle\cdot}
       + {1 \over a} \varrho v^i_{\;\,|i}
       = - {3 \over c^2} \varrho \dot U,
   \label{PN-1} \\
   & & {1 \over a} \left( a v_i \right)^{\displaystyle\cdot}
       - {1 \over a} U_{,i}
       = {1 \over c^2} {1 \over a} \left[
       2 \Phi_{,i}
       + \left( a P_i \right)^{\displaystyle\cdot}
       \right],
   \label{PN-2} \\
   & & {\Delta \over a^2} U
       + 4 \pi G \delta \varrho
       = - {1 \over c^2} \Bigg\{
       {1 \over a^2} \left[
       2 \Delta \Phi
       + \left( a P^i_{\;\;|i} \right)^{\displaystyle\cdot} \right]
   \nonumber \\
   & & \qquad
       + 3 \ddot U
       + 9 {\dot a \over a} \dot U
       + 6 {\ddot a \over a} U
       \Bigg\},
   \label{PN-3} \\
   & &
       0 = - {\Delta \over a^2} P_i
       - 16 \pi G \varrho v_i
       + {1 \over a} \left( {1 \over a} P^j_{\;\;|j}
       + 4 \dot U + 4 {\dot a \over a} U \right)_{,i}.
   \nonumber \\
   \label{PN-4}
\eea By taking a divergence, Equation (\ref{PN-4}) gives \bea
   & & 4 \pi G \varrho a v^i_{\;\,|i}
       = \Delta \left( \dot U + H U \right),
   \label{PN-4b}
\eea which follows from the 0PN order of Equations (\ref{PN-1}) and
(\ref{PN-3}). By introducing a combination \bea
   & & {\cal U}_{,i}
       \equiv U_{,i}
       + {1 \over c^2} \left[ 2 \Phi_{,i}
       + \left( a P_i \right)^{\displaystyle\cdot}
       \right],
\eea Equations (\ref{PN-2}) and (\ref{PN-3}) become \bea
   & & {1 \over a} \left( a v_i \right)^{\displaystyle\cdot}
       - {1 \over a} {\cal U}_{,i}
       = 0,
   \label{PN-2-tilde} \\
   & & {\Delta \over a^2} {\cal U}
       + 4 \pi G \delta \varrho
       = - {3 \over c^2} \left(
       \ddot U
       + 3 {\dot a \over a} \dot U
       + 2 {\ddot a \over a} U
       \right).
   \label{PN-3-tilde}
\eea

From Equations (\ref{PN-1})-(\ref{PN-3}) we can derive (Takada \& Futamase 1999) \bea
   & & \ddot \delta + 2 H \dot \delta
       - 4 \pi G \varrho \delta
       = {3 \over c^2} \left( {\dot a \over a} \dot U
       + 2 {\ddot a \over a} U \right).
   \label{PN-delta-eq1}
\eea Since terms in RHS are already 1PN order, using Equation
(\ref{PN-3}) to the Newtonian order, we have \bea
   & & \ddot \delta + 2 H \dot \delta
       - 4 \pi G \varrho \delta
       = - {12 \pi G \varrho a^2 \over c^2 \Delta}
       \left[ H \dot \delta
       + \left( 2 \dot H + H^2 \right) \delta \right].
   \nonumber \\
   \label{PN-delta-eq2}
\eea
Similarly we have \bea
   & & \!\!\!\!\!\!\!\!\!\!
       \left( a \nabla \cdot {\bf v} \right)^{\displaystyle\cdot\displaystyle\cdot}
       + H \left( a \nabla \cdot {\bf v} \right)^{\displaystyle\cdot}
       - 4 \pi G \varrho
       \left( a \nabla \cdot {\bf v} \right)
       = 0,
   \label{PN-v-eq} \\
   & & \!\!\!\!\!\!\!\!\!\!
       \left( a {\cal U} \right)^{\displaystyle\cdot\displaystyle\cdot}
       + 2 H \left( a {\cal U} \right)^{\displaystyle\cdot}
       - 4 \pi G \varrho
       \left( a {\cal U} \right)
       = 0,
   \label{PN-U-eq}
\eea where we used Equation (\ref{PN-4b}) to appropriate order.
The terms in the RHS of Equation (\ref{PN-delta-eq2}) are linear contributions from 1PN correction terms. This apparently shows that the cosmological PN approximation is valid only far inside the horizon with $c^2 \Delta
\gg 12 \pi G \varrho a^2 \sim H^2 \sim \dot H$; as we approach the horizon scale higher order PN corrections, with $({a^2 H^2 \over c^2 \Delta})^n \sim ({GM \over Rc^2})^n$ order correction factors for $n$PN order, become important.

It is remarkable that for the perturbed velocity and perturbed gravitational potential in Equations (\ref{PN-v-eq}) and (\ref{PN-U-eq}) we have {\it no} PN correction terms to the 1PN order. We have not imposed the temporal gauge condition to derive Equations (\ref{PN-delta-eq1})-(\ref{PN-v-eq}), and the variables $\delta$, ${\bf v}$, $U$ and ${\cal U}$ are gauge-invariant: this is the subject we discuss below.

In our 1PN metric convention in Equation (\ref{metric-1PN}) we already have taken spatial gauge condition by setting $g_{ij} = a^2 ( 1 + c^{-2} 2 V ) \delta_{ij}$; for a thorough examination of the gauge issue in the PN approximation see Section 6 in PN2008. Under the remaining gauge transformation $\widehat x^a = x^a + \xi^a (x^e)$ with \bea
   & & \xi^0 = {1 \over c} \xi^{(2)0} + {1 \over c^3} \xi^{(4)0},
\eea we can set $\xi^{(2)0} = 0$ without losing any generality: see Equation (173) in PN2008. Then, $U$ does not depend on the gauge and we have [see Equations (175) and (176) in PN2008] \bea
   & & \widehat P_i = P_i - {1 \over a} \xi^{(4)0}_{\;\;\;\;\;\;,i}, \quad
       \widehat \Phi = \Phi + {1 \over 2} \dot \xi^{(4)0}.
\eea
Thus, in the PN approach we have freedom to impose the temporal gauge
(hypersurface) condition on $P^i_{\;\;|i}$ or $\Phi$. A combination
$2 \Phi_{,i} + ( a P_i )^{\displaystyle\cdot}$, thus ${\cal U}$, is gauge-invariant. The fluid variables $\delta \varrho$ and ${\bf v}$ in our case of vanishing internal energy and the flux are gauge invariant: see Section 6 in PN2008.

In PN2008 we have introduced several temporal gauge conditions summarized in Table 1. In Table 1, the propagation speed indicates the speed of propagation of the gravitational potential $U$ in the Equation (\ref{PN-3}) or (\ref{PN-3-tilde}). For example, the harmonic gauge condition makes the Laplacian operator $\Delta$ in the 0PN limit replaced by a d'Alembertian operator $\Box$ by the 1PN correction terms, thus making the Poisson's equation in 0PN limit to a wave equation with the propagation speed $c$ by the 1PN correction. Similarly, the uniform-expansion gauge (and the Chandrasekhar's gauge) leaves the action-at-a-distance nature of the Poisson's equation, and the transverse-shear gauge makes Equation (\ref{PN-3}) to be no longer a wave equation. Apparently, the propagation speed of the gravitational potential depends on the gauge choice. However the propagation speed of covariantly gauge invariant Weyl tensor naturally does not depend on the gauge choice and is always $c$: see Section 7 of PN2008.

\begin{widetext}
\begin{center}
\begin{tabular}{l|l|l|l|l}
    \hline \hline
        PN Gauge & PN Definition & Propagation & PT & PT Definition \\
        & &  speed & Gauge & \\
    \hline \hline
        General gauge & ${1 \over a} P^i_{\;\;|i} + n \dot U + m H U = 0$ & $c/\sqrt{n-3}$ &
        & $ n \kappa + (n-3) {c\Delta \over a} \chi
        + 3(n-m) H \varphi \equiv 0$ \\
    \hline
        Chandrasekhar's gauge & $n=3$, $m= {\rm arbitrary}$ & $\infty$ &
        & $ \kappa + (3-m) H \varphi \equiv 0$ \\
    \hline
        Uniform-expansion gauge & $n = 3 = m$ & $\infty$ & UEG &  $\kappa \equiv 0$\\
    \hline
        Transverse-shear gauge & $n = 0 = m$ & $-ic/\sqrt{3}$ & ZSG & $\chi \equiv 0$ \\
    \hline
        Harmonic gauge & $n= 4$, $m = {\rm arbitrary}$ & $c$ &
        & 
        $\kappa + {1 \over 4} {c\Delta \over a} \chi + {3 \over 4} (4-m) H \varphi \equiv 0$ \\
    \hline \hline
\end{tabular}
\end{center}
Table 1. Comparison of often used gauge conditions in the PN approximation and in the perturbation theory (PT). The ZSG and UEG are acronyms of the zero-shear (or conformal-Newtonian, or longitudinal) gauge and the uniform-expansion (or uniform-Hubble) gauge, respectively; see Sections \ref{sec:ZSG} and \ref{sec:UEG}. The harmonic gauge in the PT is discussed in Section \ref{sec:Harmonic-gauge}. The gauge transformation properties and the gauge issue in the PN approach was thoroughly discussed in Section 6 of PN2008. The `general gauge' condition covering most of the gauge conditions introduced in the PN approach was identified in Equation (210) of PN2008; here $n$ and $m$ are arbitrary real numbers. Notice the dependence of the propagation speed of the potential on the gauge choice of $n$; see Equation (213) of PN2008. The propagation speed issue (the gauge dependence of the propagation speed of potential and the real propagation speed of gravity) is resolved based on the Weyl tensor in Section 7 of PN2008.
\end{widetext}

{\it Assuming} $\Lambda = 0 = K$, {\it for} the growing mode we have $\dot U = 0$, and \bea
   \delta = \delta^{\rm (N)} + {2 \over c^2} U, \quad
       {1 \over a} \nabla \cdot {\bf v}
       = - \dot \delta, \quad
       {\cal U} = - {4 \pi G \varrho a^2 \over \Delta} \delta^{\rm (N)},
   \nonumber \\
   \label{delta-sol-1PN}
\eea where $\delta^{\rm (N)}$ is the Newtonian solution with $\delta^{\rm (N)} \propto t^{2/3}, \; t^{-1}$; solution for $\delta$ was presented in Takada \& Futamase (1999). This implies that $2 \Phi_{,i} + (a P_i)^{\displaystyle\cdot} = 0$, thus ${\cal U} = U$.
The growing mode solutions valid to 1PN order are \bea
   & & \delta = - {2 \over 5}
       \left( {c^2 \Delta \over a^2 H^2} - 3 \right) C, \quad
       {1 \over a} \nabla \cdot {\bf v}
       = {2 \over 5} {c^2 \Delta \over a^2 H}C,
   \nonumber \\
   & &
       {1 \over c^2} {\cal U}
       = {3 \over 5} C.
   \label{sol-1PN}
\eea This should be compared with the Newtonian solutions in Equation (\ref{sol-N}). Notice that only $\delta$ has 1PN correction terms, whereas ${\bf v}$ and $U$ have {\it no} PN correction to 1PN order.

%
%
\section{Linear perturbation theory}

We consider scalar-type perturbation in a {\it flat} Robertson-Walker
background. Our metric convention is (Bardeen 1988) \bea
   & & ds^2 = - \left( 1 + 2 \alpha \right)  c^2 d t^2
       - 2 a \beta_{,i} c dt d x^i
   \nonumber \\
   & & \qquad
       + a^2 \left[ \left( 1 + 2 \varphi \right) \delta_{ij}
       + 2 \gamma_{,ij} \right] d x^i d x^j.
   \label{metric-pert}
\eea We consider a flat background with vanishing pressure and
anisotropic stress. The energy-momentum tensor is \bea
   T^0_0 = - \left( \varrho + \delta \varrho \right) c^2, \quad
       T^0_i = - a \varrho c v_{,i},
       \quad
       T^i_j = 0,
   \label{Tab-pert}
\eea where $x^0 \equiv c dt$. To the linear order, the basic perturbation equations are
(Bardeen 1988; Hwang 1991) \bea
   & & \kappa \equiv 3 H \alpha - 3 \dot \varphi
       - c {\Delta \over a^2} \chi,
   \label{eq1} \\
   & & 4 \pi G \delta \varrho + H \kappa
       + c^2 {\Delta \over a^2} \varphi
       = 0,
   \label{eq2} \\
   & & \kappa + c {\Delta \over a^2} \chi
       - {12 \pi G \over c^2} \varrho a v
       = 0,
   \label{eq3} \\
   & & {1 \over c} \dot \chi
       + {1 \over c} H \chi - \varphi - \alpha
       = 0,
   \label{eq4} \\
   & & \dot \kappa + 2 H \kappa
       - 4 \pi G \delta \varrho
       + \left( 3 \dot H + c^2 {\Delta \over a^2} \right) \alpha
       = 0,
   \label{eq5} \\
   & & \delta \dot \varrho
       + 3 H \delta \varrho
       - \varrho \left( \kappa - 3 H \alpha
       + {\Delta \over a} v \right)
       = 0,
   \label{eq6} \\
   & & {1 \over a} \left( a v \right)^{\displaystyle\cdot}
       - {c^2 \over a} \alpha
       = 0,
   \label{eq7}
\eea where $\chi \equiv a \beta + a^2 \dot \gamma/c$; $\kappa$ is the perturbed part of the trace of extrinsic curvature (minus of perturbed expansion scalar of the normal-frame vector field). These equations are presented {\it without} taking the gauge conditions.

Under the gauge transformation $\widehat x^a
= x^a + \xi^a (x^e)$ we have (Bardeen 1988; Hwang 1991) \bea
   & & \widehat \alpha = \alpha - {1 \over c} \dot \xi^0, \quad
       \widehat \delta = \delta + {3 \over c} H \xi^0, \quad
       \widehat v = v - {c \over a} \xi^0,
   \nonumber \\
   & &
       \widehat \chi = \chi - \xi^0, \quad
       \widehat \kappa = \kappa + \left( {3 \dot H \over c}
       + c {\Delta \over a^2} \right) \xi^0,
   \nonumber \\
   & &
       \widehat \varphi = \varphi - {1 \over c} H \xi^0.
   \label{GT}
\eea The gauge conditions include the hypersurface or slicing (temporal gauge)
condition and congruence (spatial gauge) condition. Our equations
are arranged using only spatially gauge invariant variables; for example, $\chi$ is a spatially gauge-invariant (being independent of $\xi^i$) combination which is the same as $\chi$ in $\gamma = 0$ spatial gauge condition (Bardeen 1988). Up to this point is a review of the linear perturbation theory.

Comparing Equations (\ref{metric-pert}) and (\ref{Tab-pert}) with Equations
(\ref{metric-1PN}) and (\ref{Tab-1PN}) we have \bea
   & &
       \alpha = - {1 \over c^2} U
       - {1 \over c^4} 2 \Phi, \quad
       \varphi = {1 \over c^2} V, \quad
       \chi_{,i} = {1 \over c^3} a P_i,
   \nonumber \\
   & &
       \kappa = - {1 \over c^2} \left( {1 \over a} P^i_{\;\;|i}
       + 3 \dot V + 3 H U \right),
   \nonumber \\
   & &
       \delta^{\rm PT} = \delta^{\rm PN}, \quad
       v_{,i} = - v_i + {1 \over c^2} P_i,
\eea where the left and the right hand sides are perturbation
variables in perturbation theory (${\rm PT}$) and in the 1PN approximation (${\rm PN}$), respectively. Notice that we have \bea
   & & v_i = - v_{\chi,i}, \quad
       {\cal U} = - c^2 \alpha_\chi,
\eea where \bea
   & & v_\chi \equiv v - {c \over a} \chi, \quad
       \alpha_\chi \equiv \alpha - {1 \over c} \dot \chi,
\eea are gauge-invariant combinations; $v_\chi$ is our notation of gauge-invariant combination made of $v$ and $\chi$ which is the same as $v$ in the zero-shear gauge $\chi \equiv 0$ (Hwang 1991). To 1PN order $\varphi$, $\delta$, $v_i$ are gauge invariant because we can set $\xi^{(2)0} \equiv 0$ in the PN approximation. Notice that gauge transformation properties of the variables in the PN approximation {\it differ} from the ones in the perturbation theory.

In terms of the ADM (Arnowitt-Deser-Misner) and the covariant notations we have (Ehlers 1961; Arnowitt et al. 1962; Ellis 1971, 1973; Bertschinger 1996)
\bea
   & & \!\!\!\!\!\!\!\!\!\!
       N = 1 - \alpha, \quad
       a_i = \alpha_{,i}, \quad
       R^{(h)} = - 4 {\Delta \over a^2} \varphi,
   \nonumber \\
   & & \!\!\!\!\!\!\!\!\!\!
       K^i_i = {1 \over c} \left( - 3 H + \kappa \right), \quad
       \overline K_{ij} = \left( \nabla_i \nabla_j - {1 \over 3} \delta_{ij} \Delta \right) \chi,
\eea where $N$, $a_c$, $R^{(h)}$, $K^i_i$, and $\overline K_{ij}$ are the lapse function, acceleration vector of the normal four-vector ($a_c \equiv n_{c;b} n^b$ with $n_i \equiv 0$), the scalar curvature of the normal hypersurface, the trace of extrinsic curvature (or expansion scalar of the normal four-vector $\theta \equiv n^a_{\;\; ;a}$ with a minus sign), and the trace-free part of extrinsic curvature (or the shear-tensor $\sigma_{ab}$ of the normal four-vector), respectively. Thus the perturbed variables $\alpha$, $\varphi$, $\kappa$ and $\chi$ can be interpreted as the perturbed part of the lapse function (or the acceleration), perturbed three-space curvature, perturbed part of the trace of extrinsic curvature (or the perturbed expansion of the normal four-vector with a minus sign), and the shear. Thus we naturally have the following names for the temporal gauge conditions: the synchronous gauge $\alpha \equiv 0$, the uniform-curvature gauge $\varphi \equiv 0$, the uniform-expansion gauge $\kappa \equiv 0$, the zero-shear gauge $\chi \equiv 0$, and the comoving gauge $v \equiv 0$.

With $\kappa$ defined in Equation (\ref{eq1}) we can show that
Equations (\ref{eq6}), (\ref{eq7}) and (\ref{eq5}) exactly reproduce
Equations (\ref{PN-1})-(\ref{PN-3}), respectively. Equation (\ref{eq3})
gives \bea
   & & 4 \pi G \varrho a v_i = \left( \dot U + H U \right)_{,i},
\eea which gives Equation (\ref{PN-4b}).

From Equations (\ref{eq5})-(\ref{eq7}) and Equation (\ref{eq7}),
respectively, we have \bea
   & & \ddot \delta + 2 H \dot \delta - 4 \pi G \varrho \delta
       = - 3 \left( {\dot a \over a} \dot \alpha + 2 {\ddot a \over a} \alpha
       \right),
   \label{delta-eq} \\
   & & {1 \over a} \left( a v \right)^{\displaystyle\cdot}
       = {c^2 \over a} \alpha.
\eea Notice that these forms of equation are valid without taking
the gauge conditions. Although Equation (\ref{delta-eq}) is in the same
form as Eq.\ (\ref{PN-delta-eq1}) with an identification $\alpha
\equiv - U/c^2$, we will show that only in certain hypersurface
conditions specified in the perturbation theory the two equations
become identical.

%
%
\section{1PN limits of linear perturbations}

\subsection{Zero-shear hypersurface}
                                              \label{sec:ZSG}

The zero-shear hypersurface condition takes $\chi \equiv 0$. From
Equation (\ref{delta-eq}), Equations (\ref{eq1}), (\ref{eq2}) and
(\ref{eq4}), and Equation (\ref{eq7}), respectively, we have \bea
   & & \ddot \delta_\chi + 2 H \dot \delta_\chi - 4 \pi G \varrho \delta_\chi
       = - 3 \left( H \dot \alpha_\chi + 2 {\ddot a \over a} \alpha_\chi \right),
   \label{delta-eq-ZSG} \\
   & & \left( c^2 {\Delta \over a^2}
       - 3 H^2 \right) \alpha_\chi
       - 3 H \dot \alpha_\chi
       = 4 \pi G \varrho \delta_\chi,
   \label{Poisson-eq-ZSG} \\
   & & {1 \over a} \left( a v_\chi \right)^{\displaystyle\cdot}
       = {c^2 \over a} \alpha_\chi.
   \label{dot-v-eq-ZSG}
\eea By identifying \bea
   & & \delta = \delta_\chi, \quad
       U \equiv - {1 \over c^2} \alpha_\chi,
\eea Equation (\ref{delta-eq-ZSG}) becomes Equation
(\ref{PN-delta-eq1}).

We can derive closed form equations for $\delta_\chi$, $v_\chi$ and $\alpha_\chi$. From Equations (\ref{eq1})-(\ref{eq7}) we have (Hwang \& Noh 1999) \bea
   & & \ddot \delta_\chi
       + 2 H \dot \delta_\chi
       - 8 \pi G \varrho \delta_\chi
       = { 3 H^2 + 6 \dot H +c^2 {\Delta \over a^2} \over
       3 H^2 - c^2 {\Delta \over a^2} \left(
       1 + {c^2 \Delta \over 12 \pi G \varrho a^2} \right)}
   \nonumber \\
   & & \qquad \times
       \left[ H \dot \delta_\chi
       + \left( 4 \pi G \varrho
       + c^2 {\Delta \over 3 a^2} \right) \delta_\chi \right],
   \label{delta-eq-exact-ZSG} \\
   & & \left( a v_\chi \right)^{\displaystyle\cdot\displaystyle\cdot}
       + H \left( a v_\chi \right)^{\displaystyle\cdot}
       - 4 \pi G \varrho
       \left( a v_\chi \right) = 0,
   \label{v-eq-ZSG} \\
   & & \left( a \alpha_\chi \right)^{\displaystyle\cdot\displaystyle\cdot}
       + 2 H \left( a \alpha_\chi \right)^{\displaystyle\cdot}
       - 4 \pi G \varrho
       \left( a \alpha_\chi \right)
       = 0.
   \label{alpha-eq-ZSG}
\eea From Equation (\ref{eq4}) we have $\varphi_\chi = - \alpha_\chi$.
In the sub-horizon limit Equation (\ref{delta-eq-exact-ZSG}) properly reproduces Equation (\ref{PN-delta-eq2}) to the 1PN limit. More remarkable is the case
of Equations (\ref{v-eq-ZSG}) and (\ref{alpha-eq-ZSG}) which {\it exactly} reproduce Newtonian equations in Equations (\ref{v-eq-N}) and (\ref{delta-Phi-eq-N}) which are valid to 1PN order: see Equations (\ref{PN-v-eq}) and (\ref{PN-U-eq}). Equations
(\ref{delta-eq-exact-ZSG})-(\ref{alpha-eq-ZSG}), are valid in
fully relativistic situation and in the general scale. The 1PN approximation corresponds to the sub-horizon limit with ${c^2 \Delta \over a^2 H^2} \gg 1$.

The growing mode {\it exact} solutions for $\Lambda = 0 = K$ are (Hwang 1994) \bea
   & & \delta_\chi
       = {2 \over 5} \left( 3 - {c^2 \Delta \over a^2 H^2} \right) C, \quad
       v_\chi
       = - {2 \over 5} {c^2 \over aH} C,
   \nonumber \\
   & &
       \varphi_\chi
       = - \alpha_\chi
       = {3 \over 5} C.
   \label{sol-ZSG}
\eea A complete set of exact solutions in the presence of $K$ and $\Lambda$ is presented in the Table 1 of Hwang (1994). Comparing with Equations (\ref{sol-N}) and (\ref{sol-1PN}), in the small-scale limit we have \bea
   & & \delta_\chi = \delta^{(\rm 1PN)}, \quad
       - \nabla v_\chi = {\bf v}^{(\rm 1PN)}, \quad
       \alpha_\chi = - \varphi_\chi
       = - {1 \over c^2} {\cal U}.
   \nonumber \\
\eea

\subsection{Uniform-expansion hypersurface}
                                              \label{sec:UEG}

The uniform-expansion hypersurface condition takes $\kappa \equiv
0$. From Equation (\ref{delta-eq}), Equation (\ref{eq5}), and Equation
(\ref{eq7}), respectively, we have \bea
   & & \ddot \delta_\kappa + 2 H \dot \delta_\kappa - 4 \pi G \varrho \delta_\kappa
       = - 3 \left( H \dot \alpha_\kappa + 2 {\ddot a \over a} \alpha_\kappa \right),
   \label{delta-eq-UEG} \\
   & & \left( c^2 {\Delta \over a^2} + 3 \dot H \right) \alpha_\kappa
       = 4 \pi G \varrho \delta_\kappa.
   \label{Poisson-eq-UEG} \\
   & & {1 \over a} \left( a v_\kappa \right)^{\displaystyle\cdot}
       = {c^2 \over a} \alpha_\kappa.
   \label{v-eq-UEG}
\eea By identifying \bea
   & & \delta = \delta_\kappa, \quad
       U \equiv - {1 \over c^2} \alpha_\kappa,
\eea Equation (\ref{delta-eq-UEG}) becomes Equation (\ref{PN-delta-eq1}).

We can derive closed form equations for $\delta_\kappa$, $v_\kappa$ and $\alpha_\kappa$. From Equations (\ref{eq1})-(\ref{eq7}) we have (Hwang \& Noh 1999)
\bea
   & & \ddot \delta_\kappa
       + 2 H \dot \delta_\kappa
       - 4 \pi G \varrho \delta_\kappa
       = {a \dot H \over H} \left( {3 H^2/a \over
       3 \dot H + c^2 {\Delta \over a^2}} \delta_\kappa \right)^{\displaystyle\cdot},
   \label{delta_kappa-eq} \\
   & & \left( a v_\kappa \right)^{\displaystyle\cdot\displaystyle\cdot}
       + H \left( a v_\kappa \right)^{\displaystyle\cdot}
       - 4 \pi G \varrho \left( a v_\kappa \right)
       = {3 \over 3 \dot H + c^2 {\Delta \over a^2}}
   \nonumber \\
   & & \qquad
       \times
       \left[
       - \left( \ddot H + 3 H \dot H \right)
       \left( a v_\kappa \right)^{\displaystyle\cdot}
       + \dot H^2 \left( a v_\kappa \right) \right],
   \label{v_kappa-eq} \\
   & & \left( a \varphi_\kappa \right)^{\displaystyle\cdot\displaystyle\cdot}
       + 2 H \left( a \varphi_\kappa \right)^{\displaystyle\cdot}
       - 4 \pi G \varrho
       \left( a \varphi_\kappa \right)
   \nonumber \\
   & & \qquad
       = {1 \over a^2 H} \left( {3 a^3 H^2 \dot H \over 3 \dot H
       + c^2 {\Delta \over a^2}} \varphi_\kappa \right)^{\displaystyle\cdot},
   \label{varphi_kappa-eq} \\
   & & \alpha_\kappa = - {c^2 {\Delta \over a^2} \over
       3 \dot H + c^2 {\Delta \over a^2}} \varphi_\kappa.
   \label{alpha-varphi_kappa}
\eea In the sub-horizon limit Equation (\ref{delta_kappa-eq}) properly reproduces Equation (\ref{PN-delta-eq2}) to the 1PN limit, whereas Equations (\ref{v_kappa-eq}) and (\ref{varphi_kappa-eq}) produce only the Newtonian equation. Equations (\ref{delta_kappa-eq})-(\ref{alpha-varphi_kappa}), however, are valid in fully relativistic situation and in the general scale. From Equations (\ref{eq2}) and (\ref{eq5}), respectively, we have \bea
   & & c^2 {\Delta \over a^2} \varphi_\kappa
       = - 4 \pi G \varrho \delta_\kappa,
   \nonumber \\
   & &
       \left( c^2 {\Delta \over a^2} + 3 \dot H \right) \alpha_\kappa
       = 4 \pi G \varrho \delta_\kappa,
\eea either of which can be regarded as the Poisson's equation.

The growing mode {\it exact} solutions for $\Lambda = 0 = K$ are (Hwang 1994) \bea
   & & \!\!\!\!\!\!\!\!\!\!
       \delta_\kappa
       = {c^2 \Delta \over c^2 \Delta - 12 \pi G \varrho a^2}
       \left( 3 - {2 \over 5} {c^2 \Delta \over a^2 H^2} \right) C,
   \nonumber \\
   & & \!\!\!\!\!\!\!\!\!\!
       v_\kappa
       = - {2 \over 5} {c^2 \over aH}
       {c^2 \Delta \over c^2 \Delta - 12 \pi G \varrho a^2} C,
   \nonumber \\
   & & \!\!\!\!\!\!\!\!\!\!
       \varphi_\kappa
       = - {4 \pi G \varrho a^2 \over c^2 \Delta} \delta_\kappa, \quad
       \alpha_\kappa
       = - {c^2 \Delta \over c^2 \Delta - 12 \pi G \varrho a^2} \varphi_\kappa.
   \label{sol-UEG}
\eea In the large-scale limit, we have $\varphi_\kappa = C$. Comparing with Equations (\ref{sol-N}) and (\ref{sol-1PN}), in the sub-horizon limit we have \bea
   & & \delta_\kappa = \delta^{(\rm 1PN)},
\eea and \bea
   & & - \nabla v_\kappa = {\bf v}^{(\rm N)}, \quad
       \alpha_\kappa
       = - \varphi_\kappa
       = - {1 \over c^2} U.
\eea only to Newtonian order.

\subsection{Comoving hypersurface}
                                              \label{sec:CG}

The comoving hypersurface condition takes $v \equiv 0$. From Equation
(\ref{eq7}) this implies $\alpha = 0$ which is the synchronous hypersurface condition. From Equations (\ref{eq5}) and (\ref{eq6}) we have \bea
   & & \ddot \delta_v + 2 H \dot \delta_v - 4 \pi G \varrho \delta_v
       = 0,
   \label{delta-eq-CG} \\
   & & \left( a^2 \kappa_v \right)^{\displaystyle\cdot\displaystyle\cdot}
       + H \left( a^2 \kappa_v \right)^{\displaystyle\cdot}
       - 4 \pi G \varrho
       \left( a^2 \kappa_v \right) = 0.
   \label{v-eq-CG}
\eea Thus, in this hypersurface condition equation for $\delta_v$ is exactly
the same as the Newtonian order equation with no PN correction. With an identification \bea
   & & \kappa_v \equiv - {1 \over a} \nabla \cdot {\bf v},
\eea equation for $\kappa_v$ is exactly the same as Newtonian one in Equation (\ref{v-eq-N}) with no PN correction. Equations (\ref{delta-eq-CG}) and (\ref{v-eq-CG}) are valid in the fully relativistic situation and in the general scale. From the gauge transformation properties in Equation (\ref{GT}) we have \bea
   \kappa_v
       \equiv \kappa + \left( {3 \dot H \over c^2} + {\Delta \over a^2} \right) a v
       \equiv \left( {3 \dot H \over c^2} + {\Delta \over a^2} \right) a v_\kappa.
\eea
Equations (\ref{eq1}), (\ref{eq3}) and (\ref{eq7}) give \bea
   & & \dot \varphi_v = 0.
   \label{varphi_v-eq}
\eea The coefficient of growing solution $C$ is introduced based on the powerful conservation property of $\varphi_v$ as  $\varphi_v = C$.

From the gauge transformation properties in Equation (\ref{GT}) we have
\bea
   & & \delta_v \equiv \delta + {3 a H \over c^2} v
       = \delta_\chi + {3 a H \over c^2} v_\chi
       = \delta_\kappa + {3 a H \over c^2} v_\kappa
   \nonumber \\
   & & \qquad
       = \delta_\varphi + {3 a H \over c^2} v_\varphi.
   \label{CG-delta-relation}
\eea Using this relation we can show that Equations (\ref{delta-eq-ZSG}), (\ref{delta-eq-UEG}) and (\ref{delta-eq-UCG}) are consistent with Eq.\ (\ref{delta-eq-CG}). In other words, by taking a combination $\delta_\chi + (3 a H / c^2) v_\chi$ and using Equations (\ref{delta-eq-ZSG}) and (\ref{v-eq-ZSG}) we have equation for $\delta_v$ in Eq.\ (\ref{delta-eq-CG}), and similarly for
$\delta_\kappa$ and $\delta_\varphi$ in Equations (\ref{delta-eq-UEG}) and (\ref{delta-eq-UCG}).

The growing mode exact solutions for $\Lambda = 0 = K$ are \bea
   \delta_v
       = - {2 \over 5} {c^2 \Delta \over a^2 H^2} C, \quad
       \kappa_v
       = - {2 \over 5} {c^2 \Delta \over a^2 H} C, \quad
       \varphi_v
       = C.
   \label{sol-CG}
\eea Comparing with Equations (\ref{sol-N}) and (\ref{sol-1PN}), in the sub-horizon limit we have \bea
   & & \delta_v = \delta^{(\rm N)}, \quad
       - \kappa_v = {1 \over a} \nabla \cdot {\bf v}^{(\rm 1PN)},
\eea to 1PN order, and \bea
   & & \alpha_v = 0, \quad
       \varphi_v
       \neq {1 \over c^2} U,
\eea even to Newtonian order.

Notice that $\delta_v$ and $\kappa_v$ exactly reproduce Newtonian density and velocity perturbations without PN correction; this is true even to the second-order perturbations (Hwang \& Noh 2004). Despite such remarkable identifications, by lacking proper correspondence with PN approximation, and lacking proper gravitational potential ($\alpha_v = 0$), it is unclear whether $\delta_v$ and $\kappa_v$ can be properly interpreted in the Newtonian context: see Section \ref{sec:correspondences}, and for our suggestion, see below Equation (\ref{correspondences}).

\subsection{Uniform-curvature hypersurface}
                                              \label{sec:UCG}

The uniform-curvature hypersurface condition takes $\varphi \equiv 0$. From Equations (\ref{eq1}), (\ref{eq3}) and (\ref{eq5})-(\ref{eq7}), and Equations (\ref{eq2}) and (\ref{eq5}), respectively, we have \bea
   & & \ddot \delta_\varphi + 2 H \dot \delta_\varphi
       - 4 \pi G \varrho \delta_\varphi
       = 3 H^2 \alpha_\varphi,
   \label{delta-eq-UCG} \\
   & & \left( c^2 {\Delta \over a^2} + 3 \dot H \right) \alpha_\varphi
       = {4 \pi G \varrho \over H} \left( \dot \delta_\varphi
       - {\dot H \over H} \delta_\varphi \right).
   \label{Poisson-eq-UCG}
\eea By combining these two equations, to the 1PN order we have \bea
   \!\!\!\!\! \ddot \delta_\varphi + 2 H \dot \delta_\varphi
       - 4 \pi G \varrho \delta_\varphi
       = {12 \pi G \varrho a^2 \over c^2 \Delta}
       \left( H \dot \delta_\varphi - \dot H \delta_\varphi \right).
   \label{delta_varphi-eq}
\eea Compared with Equation (\ref{PN-delta-eq2}) it coincides in Newtonian order but differs in 1PN order.

From Equations (\ref{eq1}), (\ref{eq3}) and (\ref{eq7}) we have $(a H v_\varphi)^\cdot = 0$ and $(a^3 H^2 \alpha)^\cdot = 0$, thus \bea
   & & v_\varphi
       \propto {H \over a \varrho} \alpha_\varphi \propto {1 \over a H},
\eea with no decaying mode. From Equations (\ref{eq2}) and (\ref{eq5}) we can derive \bea
   & & \left[ {H^2 \over 3 \dot H + c^2 {\Delta \over a^2}}
       \left( {\delta_\varphi \over H} \right)^{\displaystyle\cdot}
       \right]^{\displaystyle\cdot} = 0.
\eea

The growing mode exact solutions for $\Lambda = 0 = K$ are \bea
   & & \delta_\varphi
       = \left( 3 - {2 \over 5} {c^2 \Delta \over a^2 H^2} \right) C, \quad
       v_\varphi
       = - {c^2 \over aH} C, \quad
       \alpha_\varphi
       = - {3 \over 2} C.
   \nonumber \\
\eea Comparing with Equations (\ref{sol-N}) and (\ref{sol-1PN}), in the small-scale limit we have \bea
   & & \delta_\varphi = \delta^{(\rm N)},
\eea only to Newtonian order, and \bea
   & & - \nabla v_\varphi \neq {\bf v}^{(\rm N)}, \quad
       \alpha_\varphi
       \neq - {1 \over c^2} U,
\eea even to Newtonian order.

\subsection{Uniform-density hypersurface}

The uniform-density hypersurface condition takes $\delta \varrho \equiv 0$. This is perfectly allowed hypersurface condition in a time varying background. Although the density perturbation in this hypersurface is null by definition, physical information like physical density perturbation is not lost and is hidden in other combination of variables in this gauge like $\delta_v \equiv \delta - 3 (aH/c^2) v \equiv - 3 (aH/c^2) v_\delta$.

\subsection{Synchronous hypersurface}

The synchronous hypersurface condition takes $\alpha \equiv 0$. In the pressureless case Equation (\ref{eq7}) gives $v \propto a^{-1}$ which is the pure gauge mode: in Equation (\ref{GT}) $\widehat \alpha = 0 = \alpha$ leaves $\xi^0 = \xi^0 ({\bf x})$, and gives $v_{\rm gauge \; mode} \propto \xi^0/a \propto a^{-1}$. By setting the gauge mode $v = 0$, it becomes the same as taking the comoving gauge.

\subsection{Harmonic hypersurface}
                                      \label{sec:Harmonic-gauge}

The harmonic or de Donder hypersurface condition takes perturbed part of $g^{bc} \Gamma^a_{bc} \equiv 0$. The temporal component gives $\dot \alpha + H \alpha + \kappa = 0$ which is quite a bad choice leaving two remnant gauge modes: see the Appendix in Hwang (1993). In the PN approximation this becomes ${1 \over a} P^i_{\;\;|i} + 4 \dot U + 4 HU = 0$ with the well known propagation speed $c$; however, we note that as presented in Table 1 the propagation speed of potential depends on the PN gauge choice. In Table 1 we presented a combination of perturbation variables which reproduces this PN definition of the harmonic gauge with a general $H U$ term.

%
%
\section{Post-Newtonian/perturbation theory correspondence}
                                    \label{sec:correspondences}

We summarize correspondences between the PN approximation and the perturbation theory in Table 2.

\begin{widetext}
\begin{center}
\begin{tabular}{l|l|l|l|l}
    \hline \hline
        \hskip 3cm & ZSG & UEG & CG \qquad\quad\quad & UCG \qquad\qquad \\
    \hline \hline
        $\delta$ \hskip .51cm 1PN & $\delta_\chi$ & $\delta_\kappa$ &  &  \\
    \hline
        \hskip .81cm 0PN Exact & & & $\delta_v$ \quad [$2^{\rm nd}$] &  \\
    \hline
        \hskip .81cm 0PN Sub-horizon \quad & $\delta_\chi$ \hskip 1.7cm [$2^{\rm nd}$] & $\delta_\kappa$ \hskip 1.7cm [$2^{\rm nd}$] & $\delta_v$ \quad [$2^{\rm nd}$] & $\delta_\varphi$ \quad [$2^{\rm nd}$] \\
    \hline \hline
        ${\bf v}$ \hskip .48cm 1PN(=0PN) & $v_\chi$ &  & $\kappa_v$ \quad [$2^{\rm nd}$] &  \\
    \hline
        \hskip .82cm 0PN Exact & $v_\chi$ & & $\kappa_v$ \quad [$2^{\rm nd}$] &  \\
    \hline
        \hskip .82cm 0PN Sub-horizon & $v_\chi$ \hskip 1.7cm [$2^{\rm nd}$] & $v_\kappa$ \hskip 1.7cm [$2^{\rm nd}$] & $\kappa_v$ \quad [$2^{\rm nd}$] &  \\
    \hline \hline
        $U$ \hskip .43cm 1PN(=0PN) & $\alpha_\chi (= - \varphi_\chi)$ &  &  &  \\
    \hline
        \hskip .83cm 0PN Exact & $\alpha_\chi (= - \varphi_\chi)$ &  &  &  \\
    \hline
        \hskip .83cm 0PN Sub-horizon & $\alpha_\chi (= - \varphi_\chi)$ \quad [$2^{\rm nd}$] & $\alpha_\kappa (= - \varphi_\kappa)$ \quad [$2^{\rm nd}$] &  &  \\
    \hline \hline
\end{tabular}
\end{center}
Table 2. The CG and UCG are acronyms of the comoving gauge and the uniform-curvature (or flat) gauge, respectively; see Sections \ref{sec:CG} and \ref{sec:UCG}. A notation [$2^{\rm nd}$] means that the Newtonian correspondence is valid to second-order perturbation in the perturbation theory (Hwang et al. 2012).
\end{widetext}

As the 1PN correction terms have ${a^2 H^2 \over c^2 \Delta}$ order factor smaller than 0PN terms, those become important as the scale approaches the horizon. Thus the PN approximation is applicable only in the sub-horizon scale. However, the Newton's theory is free from the presence of the horizon scale. This is why we have divided 0PN case into the exact (all scales) and the sub-horizon cases in Table 2.

Our comparison of PN approximation and perturbation theory shows that the zero-shear gauge is distinguished by the fact that all perturbation variables of the density, velocity and gravitational potential properly correspond to 1PN (thus, valid in the sub-horizon scale) order variables. Thus \bea
   & & \delta_\chi = \delta^{\rm (1PN)}, \quad
       - \nabla v_\chi = {\bf v}^{\rm (1PN)} = {\bf v}^{\rm (N)},
   \nonumber \\
   & &
       \alpha_\chi = - \varphi_\chi = - {1 \over c^2} {\cal U}
       = - {1 \over c^2} U.
\eea Notice that for the perturbed velocity and gravitational potential the Newtonian (0PN) equations remain valid to 1PN order.

In the uniform-expansion gauge, only the density perturbation has 1PN correspondence whereas the velocity and gravitational potential perturbations have 0PN correspondences only in the sub-horizon limit.

Newtonian correspondences of all perturbation variables (density, velocity and gravitational potential) on the sub-horizon scale in the zero-shear gauge and the uniform expansion gauge were known to the linear order (Peebles 1980; Hwang \& Noh 1999); the same correspondences are shown to be valid even to the second-order perturbations (Hwang et al. 2012).

Newtonian correspondences of linear density perturbations in various gauge conditions in the sub-horizon scale were pointed out in Bardeen (1980); the same correspondences are shown to be valid even to the second-order perturbations (Hwang et al. 2012). Only in the comoving gauge (to the linear order the same as the synchronous gauge without the gauge mode) the density perturbation has an {\it exact} correspondence to the Newtonian one (Lifshitz 1946; Bonnor 1957; Nariai 1969; Bardeen 1980).

In the case of the comoving gauge the relativistic/Newtonian correspondence is quite distinguishing: up to the second order in perturbations the perturbed density and velocity show {\it exact} (thus, valid in all scales) correspondences with the Newtonian ones (Noh \& Hwang 2004; Hwang et al. 2012); these correspondences are possible only in the comoving gauge. However, the comoving gauge implies vanishing $\alpha_v$, thus vanishing gravitational potential to the linear order.
Without the proper gravitational potential, despite the striking correspondences of density and velocity, whether proper Newtonian interpretation will be available in this gauge condition is unclear; for our suggestion, see below.

For the growing modes, Equations (\ref{sol-1PN}), (\ref{sol-ZSG}) and (\ref{sol-CG}) show that (Takada \& Futamase 1999) \bea
   \delta_\chi = \delta_v + {2 \over c^2} U, \quad
       \delta_\chi = \delta^{\rm (1PN)}, \quad
       \delta_v = \delta^{\rm (N)}.
\eea
Takada \& Futamase (1999) have identified $\delta_v$ as the physical density perturbation: see below their Equation (A18). In perturbation theory, this solution simply follows from the relations between the gauge-invariant combinations in Equation (\ref{CG-delta-relation}): $\delta_\chi = \delta_v - 3 (a H / c^2) v_\chi$. From Equations (\ref{eq2})-(\ref{eq7}) we can derive
\bea
   & & \dot \delta_v = \kappa_v,
   \label{E-conserv-E} \\
   & & \dot \kappa_v + 2 H \kappa_v = 4 \pi G \varrho \delta_v,
   \label{Mom-conserv-E} \\
   & & c^2 {\Delta \over a^2} \alpha_\chi
       = 4 \pi G \varrho \delta_v,
   \label{Poisson-E}
\eea These equations can be compared with Equations (\ref{Newtonian-eq1})-(\ref{Newtonian-eq3}) in the Newtonian context. Equation (\ref{Poisson-E}) is the well known Poisson's equations relating gauge-invariant combinations based on different gauges: density perturbation based on the comoving gauge and gravitational potential based on the zero-shear gauge; this was presented in Equations (4.3) and (4.4) of Bardeen (1980). Thus, to the linear order we may {\it identify} \bea
   \delta_v \equiv \delta, \quad
       \kappa_v \equiv - {1 \over a} \nabla \cdot {\bf u}, \quad
       \alpha_\chi \equiv - {1 \over c^2} U.
   \label{correspondences}
\eea It is likely that the proper Newtonian correspondence is available based on such mixed use of the gauges in constructing gauge-invariant combinations. Bardeen (1980) in his Equations (4.5) and (4.8) has identified $v_\chi$ instead of $\kappa_v$ as the velocity perturbation variable; as summarized in Table 2, both $v_\chi$ and $\kappa_v$ show distinguished 0PN and 1PN correspondences, but $\kappa_v$ is better in having the {\it exact} Newtonian correspondence even to the second order in all scales.

%
%
\acknowledgments

J.H.\ was supported by KRF Grant funded by the Korean Government
(KRF-2008-341-C00022).
H.N.\ was supported by grant No.\ 2010-0000302 from KOSEF funded by the Korean Government (MEST).

%
%



\begin{thebibliography}{21}
\expandafter\ifx\csname natexlab\endcsname\relax\def\natexlab#1{#1}\fi
\bibitem{ADM}
         Arnowitt, R., Deser, S., \& Misner, C. W. 1962 Gravitation: an
                introduction to current research, edited by  Witten, L.
                (Wiley, New York) 227
\bibitem{Asada-Futamase-1997}
         Asada, H., \& Futamase, T. 1997, Prog. Theor. Phys. Suppl., {128}, 123
\bibitem{Bardeen-1980}
         Bardeen, J. M. 1980, Phys. Rev. D, 22, 1882
\bibitem{Bardeen-1988}
         Bardeen, J. M. 1988, Particle Physics and Cosmology, edited by
                       Fang L., \& Zee A. (Gordon and Breach, London) 1
\bibitem{Bertschinger-1996}
         Bertschinger, E. 1996, Cosmology and Large Scale Structure, proc. Les Houches
         Summer School, Session LX, edited by Schaeffer, R., Silk, J., Spiro, M., \&
         Zinn-Justin, J. (Elsevier Science, Amsterdam) 273
\bibitem{Bertschinger-Hamilton-1994}
         Bertschinger, E., \& Hamilton, A. J. S. 1994, ApJ, {435}, 1
\bibitem{Bonnor-1957}
         Bonnor, W. B. 1957, MNRAS, 107, 104
\bibitem{Chandrasekhar-1965}
         Chandrasekhar, S. 1965, ApJ, {142}, 1488
\bibitem{Chandrasekhar-Nutku-1969}
         Chandrasekhar, S., \& Nutku, Y. 1969, ApJ, {158}, 55
\bibitem{Ehlers-1961}
         Ehlers, J. 1961, Proc. Mainz Acad. Sci. Lit., No. 11, 792; English translation 1993, Gen. Rel. Grav., 25, 1225
\bibitem{Ellis-1971}
         Ellis, G. F. R. 1971, General Relativity and Cosmology, edited by
         Sachs, R. K. (Academic Press, New York) 104
\bibitem{Ellis-1973}
         Ellis, G. F. R. 1973, Cargese Lectures in Physics, edited by
         Schatzmann, E. (Gorden and Breach, New York) 1
\bibitem{Futamase-1991}
         Futamase, T. 1988, Phys. Rev. Lett., {61}, 2175
\bibitem{Futamase-1989}
         Futamase, T. 1989, MNRAS, {237}, 187
\bibitem{Futamase-1993a}
         Futamase, T. 1993a, Prog. Theor. Phys., {86}, 389
\bibitem{Futamase-1993b}
         Futamase, T. 1993b, Prog. Theor. Phys., {89}, 581
\bibitem{Harrison-1967}
         Harrison, E. R. 1967, Rev. Mod. Phys., {39}, 862
\bibitem{H-1991}
         Hwang, J. 1991, ApJ, {375}, 443
\bibitem{IF-1993}
         Hwang, J. 1993, ApJ, {415}, 486
\bibitem{MDE-1994}
         Hwang, J. 1994, ApJ, {427}, 533
\bibitem{HN-Newtonian-1999}
         Hwang, J., \& Noh, H. 1999, Gen. Rel. Grav., {31}, 1131
\bibitem{HNG-2012}
         Hwang, J., Noh, H., \& Gong, J. 2012, ApJ, 752, 50
\bibitem{PN-2008}
         Hwang, J., Noh, H., \& Puetzfeld, D. 2008, JCAP, {03}, 010 (PN2008)
\bibitem{Jeong_etal-2011}
         Jeong, D., Gong, J., Noh, H., \& Hwang, J. 2011, ApJ, 722, 1
\bibitem{Kodama-Sasaki-1991}
         Kodama, H., \& Sasaki, M. 1984, Prog. Theor. Phys. Suppl., {78}, 1
\bibitem{Kofman-Pogosyan-1995}
         Kofman, L., \& Pogosyan, D. 1995, ApJ, {442}, 30
\bibitem{Layzer-1954}
         Layzer, D. 1954, AJ, 59, 268
\bibitem{Lemons-1988}
         Lemons, D. S. 1988, Am. J. Phys., 56, 502
\bibitem{Lifshitz-1946}
         Lifshitz, E. M. 1946, J. Phys. (USSR), {10}, 116
\bibitem{Ma-Bertschinger-1995}
         Ma, C.-P., \& Bertschinger, E. 1995, ApJ, {455}, 7
\bibitem{Mukhanov-etal-1992}
         Mukhanov, V. F., Feldman, H. A., \& Brandenberger, R. H. 1992, Phys. Rep.
         215, 203
\bibitem{Nariai-1969}
         Nariai, H. 1969, Prog. Theor. Phys., {41} 686
\bibitem{second-order-2004}
         Noh, H., \& Hwang, J. 2004, Phys. Rev. D, 69, 104011
\bibitem{Peebles-1980}
         Peebles, P. J. E. 1980, The large-scale structure of the universe,
                (Princeton Univ. Press, Princeton)
\bibitem{Shibata-Asada-1995}
         Shibata, M., \& Asada, H. 1995, Prog. Theor. Phys., 94, 11
\bibitem{Takada-Futamase-1999}
         Takada, M., \& Futamase, T. 1999, MNRAS, {306}, 64
\bibitem{Tomita-1988}
         Tomita, K. 1988, Prog. Theor. Phys., {79}, 258
\bibitem{Tomita-1991}
         Tomita, K. 1991, Prog. Theor. Phys., {85}, 1041
\end{thebibliography}
\end{document}